\documentclass[conference]{IEEEtran}

\usepackage{graphicx}
\usepackage{soul}
\usepackage[inkscapelatex=false]{svg}
\usepackage{listings}
\usepackage{comment}
\usepackage{subcaption}
\usepackage{caption}
\usepackage{float}
\usepackage{multirow}
\usepackage{array}
\usepackage{booktabs}
\usepackage[table,xcdraw,dvipsnames]{xcolor}
\usepackage{tabularx}
\usepackage{tikz}
\usepackage{amssymb}
\usepackage[most]{tcolorbox}
\usepackage[hidelinks]{hyperref}
\usepackage{enumitem}
\usepackage{caption}
\usepackage[T1]{fontenc}
\usetikzlibrary{tikzmark}

\usetikzlibrary{calc, positioning, patterns}
\usetikzlibrary{decorations.pathreplacing}
\definecolor{matchblue}{RGB}{200, 220, 255}
\definecolor{ignoregray}{RGB}{240, 240, 240}

\newcolumntype{Y}{>{\centering\arraybackslash}X}
\newcolumntype{R}[1]{>{\centering\arraybackslash}m{#1}}

\definecolor{MyWorkColor}{rgb}{0.96, 0.97, 0.99} 
\definecolor{HeaderGray}{gray}{0.85} 

\definecolor{jsonkey}{RGB}{127,0,85}
\definecolor{jsonstring}{RGB}{42,0,255}
\definecolor{jsonbg}{RGB}{245,245,245}

\newtcbox{\jsoninline}{
    on line,
    boxrule=0.5pt,
    arc=2pt,
    outer arc=2pt,
    colback=jsonbg,
    colframe=gray!60,
    boxsep=0pt,
    left=2pt,
    right=2pt,
    top=1pt,
    bottom=1pt,
    fontupper=\ttfamily\scriptsize
}

\definecolor{snortKeyword}{RGB}{0,0,180}
\definecolor{snortOption}{RGB}{180,0,0}
\definecolor{snortString}{RGB}{0,140,0}
\definecolor{snortComment}{RGB}{120,120,120}

\lstdefinelanguage{Snort}{
  morekeywords={
    alert,log,pass,drop,reject,
    tcp,udp,icmp,ip,
    msg,content,offset,depth,sid,rev,classtype,priority
  },
  sensitive=true,
  morecomment=[l]{\#},
  morestring=[b]"
}

\lstdefinestyle{snort}{
  language=Snort,
  basicstyle=\ttfamily\small,
  keywordstyle=\color{snortKeyword}\bfseries,
  identifierstyle=\color{black},
  stringstyle=\color{snortString},
  commentstyle=\color{snortComment},
  frame=single,
  rulecolor=\color{black},
  breaklines=true,
  columns=fullflexible,
  keepspaces=true,
  showstringspaces=false,
  tabsize=2
}

\definecolor{keycolor}{RGB}{28, 64, 114}    
\definecolor{stringcolor}{RGB}{45, 120, 45} 
\definecolor{numbercolor}{RGB}{180, 90, 0}  

\lstdefinelanguage{json}{
    basicstyle=\small\ttfamily,
    showstringspaces=false,
    breaklines=true,
    frame=lines,
    backgroundcolor=\color{gray!2},
    escapechar=|, 
    stringstyle=\color{stringcolor},
    string=[s]{"}{"},
    literate=
     *{:}{{{\color{black}{:}}}}{1}
      {,}{{{\color{black}{,}}}}{1}
      {\{}{{{\color{black}{\{}}}}{1}
      {\}}{{{\color{black}{\}}}}}{1}
      {[}{{{\color{black}{[}}}}{1}
      {]}{{{\color{black}{]}}}}{1},
}

\lstdefinestyle{ir_style}{
    basicstyle=\ttfamily\small, 
    breaklines=true,            
    frame=single,               
    rulecolor=\color{black!30}, 
    backgroundcolor=\color{gray!5}, 
    showstringspaces=false      
}

\definecolor{dkgreen}{rgb}{0,0.6,0}
\definecolor{gray}{rgb}{0.5,0.5,0.5}
\definecolor{mauve}{rgb}{0.58,0,0.82}
\definecolor{gray}{rgb}{0.4,0.4,0.4}
\definecolor{darkblue}{rgb}{0.0,0.0,0.6}
\definecolor{lightblue}{rgb}{0.0,0.0,0.9}
\definecolor{cyan}{rgb}{0.0,0.6,0.6}
\definecolor{darkred}{rgb}{0.6,0.0,0.0}

\lstdefinestyle{topofpage}{
  float=tp,
  floatplacement=tbp,
}

\lstdefinelanguage{xml}{
  moredelim=**[is][\color{black}]{@}{@},
  morestring=[s][\color{mauve}]{"}{"},
  commentstyle=\color{dkgreen},
  breaklines=true,
  morecomment=[s]{<!--}{-->},
  stringstyle=\color{black},
  identifierstyle=\color{lightblue},
  keywordstyle=\color{red},
  morekeywords={name, height, width, die, has_prog_routing, type, layer_offset, pattern}
}

\lstdefinelanguage{json}{
  moredelim=**[is][\color{black}]{@}{@},
  morestring=[s][\color{mauve}]{"}{"},
  commentstyle=\color{dkgreen},
  breaklines=true,
  morecomment=[s]{<!--}{-->},
  stringstyle=\color{stringcolor},
  identifierstyle=\color{lightblue},
  keywordstyle=\color{red},
  escapechar=|,
  literate=
     *{:}{{{\color{black}{:}}}}{1}
      {,}{{{\color{black}{,}}}}{1}
      {\{}{{{\color{black}{\{}}}}{1}
      {\}}{{{\color{black}{\}}}}}{1}
      {[}{{{\color{black}{[}}}}{1}
      {]}{{{\color{black}{]}}}}{1},
}

\definecolor{codegreen}{rgb}{0,0.6,0}
\definecolor{codegray}{rgb}{0.5,0.5,0.5}
\definecolor{codepurple}{rgb}{0.58,0,0.82}
\definecolor{backcolour}{rgb}{0.95,0.95,0.92}

\lstnewenvironment{algorithm}[1][]{
    \lstset{
        backgroundcolor=\color{backcolour},  
        basicstyle=\ttfamily\footnotesize,
        breaklines=true,
        mathescape=true,
        frame=tblr,
        numbers=left,
        numberstyle=\tiny,
        commentstyle=\color{codegreen},
        keywordstyle=\color{magenta},
        comment=[l]{//},
        keywords={,function, in, if, else, foreach, while, do, done, not, begin, end, continue, for, and, }
        numbers=left,
        xleftmargin=.025\textwidth,
        tabsize=2
    }
}{}

\begin{document}

\title{From Patterns to Parsers: Automatic Generation of Efficient Hardware Parsers for FPGAs}
\IEEEoverridecommandlockouts
\author{
\thanks{This research, including the development and evaluation of the tool, was conducted in a personal capacity and by the University of Waterloo. The views, findings, and conclusions expressed in this paper are solely those of the authors and do not represent the official policy or position of Meta.}
\IEEEauthorblockN{
    Tushar Garg\textsuperscript{*} and 
    Andrew Boutros\textsuperscript{\textdagger}}
\IEEEauthorblockA{
\textsuperscript{*}Meta Platforms, Inc.\\
\textsuperscript{\textdagger}Department of Electrical and Computer Engineering, University of Waterloo \\
Email: gtushar@meta.com, andrew.boutros@uwaterloo.ca}
}

\maketitle

\begin{abstract}
This work presents an open-source tool for automatically generating efficient hardware parsers from high-level specifications. 
It uses a parsing intermediate representation (PIR) that decouples application-specific frontends from a common register-transfer level (RTL) generation backend. 
The backend produces optimized, human-readable SystemVerilog, handling FSM generation, byte-alignment, and multi-cycle field straddling for arbitrary datapath widths.
The tool also extends pattern matching beyond simple equality checks by introducing custom symbolic tokens to support operations that existing parser generators cannot express, such as range validation, negation, and comparisons against external ports. 
We demonstrate two end-to-end flows using a P4 frontend for Ethernet protocol parsing and a Snort frontend for network intrusion detection, both using the same unmodified backend. 
The generated Ethernet parsers achieve up to 226\% higher operating frequency and up to 97\% fewer FPGA logic resources than prior work. 
A controlled synthetic study further shows that the tool's hierarchical pattern decomposition yields up to 8$\times$ resource utilization reduction over monolithic designs. 
Our open-source framework enables designers to rapidly implement high-performance, resource-efficient, vendor-agnostic hardware parsers for diverse applications.
\end{abstract}

\IEEEpeerreviewmaketitle

\hyphenpenalty=10000

\fontsize{10.3pt}{12.05pt}\selectfont

\section{Introduction}
\vspace{-2pt}
\label{sec:intro}

Line-rate data parsing is a fundamental operation in modern high-performance computing. As data rates outpace general-purpose processor capabilities, dedicated packet processors on FPGAs~\cite{firestone2018azure, boo2023f4t} or ASICs~\cite{bosshart2013forwarding, burstein2021nvidia} have become essential to meet throughput and latency targets.
High-performance hardware parsing circuitry is central to all these accelerators.
For instance, financial trading systems use FPGA-based parsers to decode market data with sub-microsecond latency, where parsing latency directly affects trading outcomes~\cite{lockwood2012low, boutros2017build}.
Similarly, modern smart network interface cards (SmartNICs) implement security features that require parsing and inspecting packets in real-time without throttling bandwidth~\cite{zhao2020achieving}.

Implementing efficient hardware parsers requires manually crafting complex finite state machines (FSMs) for state tracking, byte alignment, and multi-cycle field straddling across arbitrary datapath widths, a process that is slow, error-prone, and brittle under evolving protocol specifications or datapath scaling. 
Tools generating hardware from domain-specific languages, such as P4~\cite{bosshart2014p4}, have been proposed to automate this process. 
However, most commercial offerings produce encrypted, vendor-locked RTL~\cite{amd2025vitisp4}, while open-source alternatives either rely on high-level synthesis (HLS) resulting in suboptimal implementation quality or have stagnated with deprecated dependencies~\cite{gibb2013design,benacek2016p4}.

This work introduces a versatile open-source tool that bridges design productivity and implementation quality by automatically generating optimized RTL hardware parsers.
Built on our parsing intermediate representation (PIR), the tool decouples domain-specific frontend(s) from the RTL generation backend, abstracting parsing specifications into a generalized pattern matching problem that automatically handles complexities such as dynamic byte alignment and multi-cycle fields.
The tool also enables users to include symbolic tokens in their parsing specifications, which extends pattern matching beyond equality to support operations that no existing hardware parser generator can express natively, such as range validation, negation, and comparisons against external ports.
The backend is target-agnostic and template-driven using Jinja2, enabling output in plain SystemVerilog.
The generated designs are pipelined to achieve high frequencies yet remain structured and human-readable for further manual optimization, if needed.
To showcase the tool, we implement two example frontends: a P4-based frontend for Ethernet/IP protocol parsing and a Snort frontend for network security rule parsing.
In summary, our contributions in this work are:
\begin{itemize}[leftmargin=*, itemindent=0pt, labelsep=3pt, align=left]
  \item A versatile open-source\footnote{Source code link: \url{https://github.com/boutros-lab/patterns-to-parsers}} tool for automatic hardware parser generation, using a parsing intermediate representation that decouples application frontends from the RTL backend.
  \item Extended pattern matching with custom symbolic tokens for range validation, negation, and external port comparisons.
  \item Validation across two distinct domains (Ethernet protocol parsing and network intrusion detection) using the same unmodified backend, achieving up to 226\% higher operating frequency and up to 97\% and 71\% fewer lookup tables (LUTs) and flip-flops (FF) compared to prior work.
\end{itemize}
\noindent

\section{Background \& Related Work}
\vspace{-2pt}
\label{sec:background}
  
Packet parsing involves two key stages: \emph{pattern matching} to identify fields of interest, and \emph{data extraction} to retrieve and align them for downstream processing.
Strict sequential dependencies between nested headers, where parsing one header depends on fields from the previous, require complex control FSMs.
Header extraction is also resource-intensive, requiring large barrel shifters with additional complexity when fields straddle multiple data flits.

One hardware parser implementation approach~\cite{attig2011400,benacek2016p4, mashreghi2022templated} chains identical processing elements, each handling a single header and passing metadata along with the data flit to the next.
While this handles arbitrary protocol stacks, every stage must incorporate full barrel shifters for byte alignment regardless of the limited shifting actually required, resulting in significant resource overhead.
Parsing latency also scales linearly with header depth.
In contrast, our tool generates flat custom parsers implementing only the minimal required data extraction multiplexing, yielding significant latency reduction and resource savings.

Some implementations use ternary content addressable memories (TCAMs), where parsing state machines are stored in runtime-configurable tables~\cite{kozanitis2010leaping, gibb2013design, bosshart2013forwarding, kumarasamy2024fpga}. While TCAMs enable parallel lookup and can reduce identification latency, they introduce significant area overhead and power consumption.
These approaches also typically rely on speculative field extraction (i.e., extracting multiple potential header fields before identification completes) which can increase design complexity.

Other works rely on domain-specific compilers to automate parser generation from high-level descriptions.
The authors of~\cite{santiago2018p4} compile P4 into C++ and use AMD Vivado HLS to generate hardware.
While this simplifies the design process, the semantic gap between P4's packet processing model and C++'s sequential execution model forces insertion of synchronization logic, producing parsers with higher latency than hand-optimized counterparts due to conservative HLS scheduling.
Similarly, P4FPGA~\cite{wang2017p4fpga} is an open-source P4-to-FPGA compiler generating Bluespec SystemVerilog implementations, but it is limited to P4 inputs and has not seen active development for a decade.
The Xilinx SDNet commercial tool~\cite{ibanez2019p4} translates P4 directly into RTL, but produces encrypted vendor-specific IP that hinders platform migration and prevents use-case-specific optimizations.

Unlike these approaches, our tool is not bound to a specific input language and does not sacrifice implementation quality by using generic HLS tools. 
It implements a fully decoupled RTL generation backend consuming a generic intermediate representation for pattern matching and data extraction. 
The generated RTL is structured, human-readable, and vendor-agnostic, enabling designers to inspect and optimize the output for their target platform as needed.
\section{Tool Overview}
\vspace{-2pt}
\label{sec:tool_overview}

\begin{figure}[t!]
    \centering
    \includegraphics[width=1\linewidth]{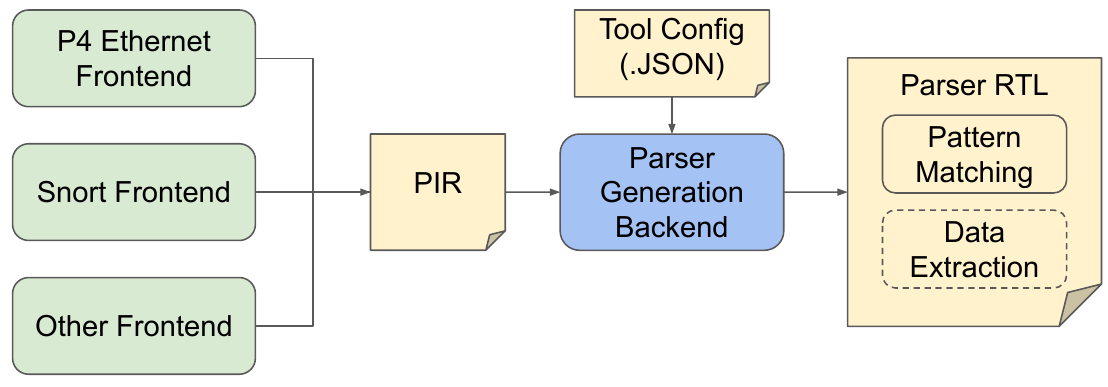}
    \caption{High-level overview of our hardware parser generation toolflow.}
    \label{fig:tool_overview}
    \vspace{-10pt}
\end{figure}

Fig.~\ref{fig:tool_overview} illustrates a high-level overview of our parser generation toolflow.
The backend translates our generic PIR to an optimized RTL implementation based on provided configuration settings, while different application-specific frontends generate parsing specifications in PIR format.
In addition, the backend also generates a SystemVerilog testbench code to validate the pattern matching and data extraction functionality using automatically generated synthetic input patterns.
This section introduces the PIR format, configuration settings, and supported parsing features.
We also implement two example frontends detailed in Sections~\ref{sec:eth} and~\ref{sec:snort}. 

\vspace{-4pt}
\subsection{Parsing Intermediate Representation}

Our PIR defines the parsing specification as a set of pattern strings along with their corresponding unique identifiers and (optional) data extraction requirements.
Pattern strings are represented at 4-bit (nibble) granularity, chosen to align with the hexadecimal representation commonly used in networking protocol specifications and tools (e.g., \texttt{tcpdump} and Wireshark).
The PIR supports three character classes:
\begin{itemize}[leftmargin=*, itemindent=0pt, labelsep=3pt, align=left]
    \item \emph{Hexadecimal Digits} (\texttt{0-9, A-F}) that define required constant matches at the corresponding nibble positions.
    \item \emph{Wildcards} (\texttt{x}) that indicate ``do not care'' positions masked during matching.
    \item \emph{Custom Symbolic Tokens (any other character):} that enable complex comparisons beyond simple equality checks, such as comparisons against external input ports, inequalities, range checks, and set membership. Their meaning is defined in the tool configuration file discussed in the next subsection.
\end{itemize}

Each pattern string is coupled with a unique output identifier, specifying the value produced when the pattern is detected in the input data stream.
Optionally, the PIR specifies data fields to extract when a pattern is matched, represented as tuples of field ID, start position, and end position.
The backend uses these to implement minimal multiplexing for extracting specified fields across all pattern strings.
If no extraction fields are specified, the generated parser implements only pattern matching.
PIR patterns are prioritized by specification order; when input data matches multiple patterns, only the first matching pattern identifier is produced.
This priority-based matching guarantees deterministic output for overlapping patterns and enables hierarchical matching where specific patterns take precedence over general fallbacks.

\begin{figure}[t!]
\centering
\begin{subfigure}[b]{\linewidth}
    \centering
    \resizebox{\linewidth}{!}{\begin{tikzpicture}[
    cell/.style={rectangle, draw, minimum width=0.7cm, minimum height=0.7cm, font=\ttfamily\large},
    desc_box/.style={
        rectangle, 
        draw, 
        minimum height=0.7cm, 
        text width=3.5cm,   
        align=center,       
        font=\sffamily\small
    },
    bit_label/.style={font=\scriptsize\sffamily, color=gray!80!black},
]

\def\rowsep{-1}  

\node[cell, fill=ignoregray] (c0) at (0,0) {x};
\node[cell, fill=matchblue]  (c1) at (0.7,0) {0};
\node[cell, fill=matchblue]  (c2) at (1.4,0) {8};
\node[cell, fill=ignoregray] (c3) at (2.1,0) {x};
\node[cell, fill=ignoregray] (c4) at (2.8,0) {x};
\node[cell, fill=ignoregray] (c5) at (3.5,0) {x};
\node[cell, fill=matchblue]  (c6) at (4.2,0) {6};

\node[cell, fill=green!20]   (c7) at (5.1,0) {1};

\node[desc_box, right=0.2cm of c7,  fill=red!20] (c8) {field1:16:23;field2:0:7};

\node[cell, fill=matchblue]  (d0) at (0,\rowsep) {1};
\node[cell, fill=matchblue]  (d1) at (0.7,\rowsep) {F};
\node[cell, fill=ignoregray] (d2) at (1.4,\rowsep) {x};
\node[cell, fill=ignoregray] (d3) at (2.1,\rowsep) {x};
\node[cell, fill=matchblue]  (d4) at (2.8,\rowsep) {A};
\node[cell, fill=ignoregray] (d5) at (3.5,\rowsep) {x};
\node[cell, fill=ignoregray] (d6) at (4.2,\rowsep) {x};

\node[cell, fill=green!20]   (d7) at (5.1,\rowsep) {2};

\node[desc_box, right=0.2cm of d7, fill=red!20] (d8) {field2:0:7};

\node[above=0.05cm of c0, bit_label] {27:24};
\node[above=0.05cm of c6, bit_label] {3:0};

\draw[<-, thick] (c0.north west) -- ++(0, 0.6) node[above, font=\sffamily\small] {MSB};
\draw[<-, thick] (c6.north east) -- ++(0, 0.6) node[above, font=\sffamily\small] {LSB};

\node[anchor=east, font=\sffamily\small] at (c0.west) {Pattern 1};
\node[anchor=east, font=\sffamily\small] at (d0.west) {Pattern 2};

\end{tikzpicture} }
    \caption{}
    \label{subfig:bitmap1}
\end{subfigure}
\par\bigskip
\begin{subfigure}[b]{\linewidth}
    \centering
    \includegraphics[width=\linewidth]{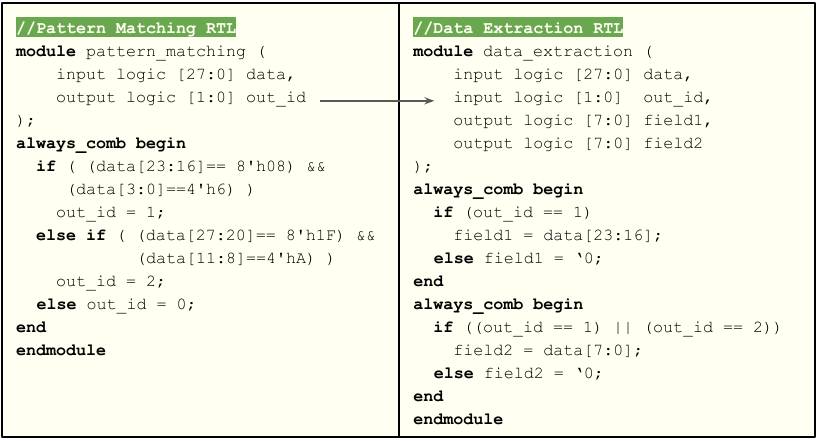}
    \caption{}
    \label{subfig:sv_pdf1}
\end{subfigure}
\caption{
(a) Bit map of two PIR entries with their output identifiers and data extraction requirements shown in (\textcolor{green!20}{\rule{1.2ex}{1.2ex}}) and (\textcolor{red!20}{\rule{1.2ex}{1.2ex}}), respectively.
(b) Generated RTL implementation for pattern matching and data extraction.
The tool generates equality comparators for 
(\textcolor{matchblue}{\rule{1.2ex}{1.2ex}}) regions and ignores (\textcolor{ignoregray}{\rule{1.2ex}{1.2ex}}) regions in the pattern matching module. The data extraction module steers the specified fields to the corresponding ports based on the pattern matching output.
}
\label{fig:bit_mapping_constant}
\vspace{-10pt}
\end{figure}
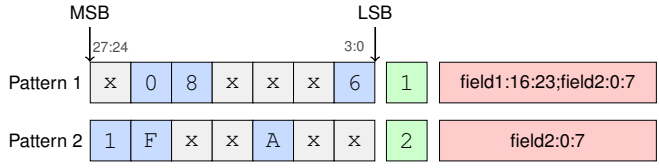
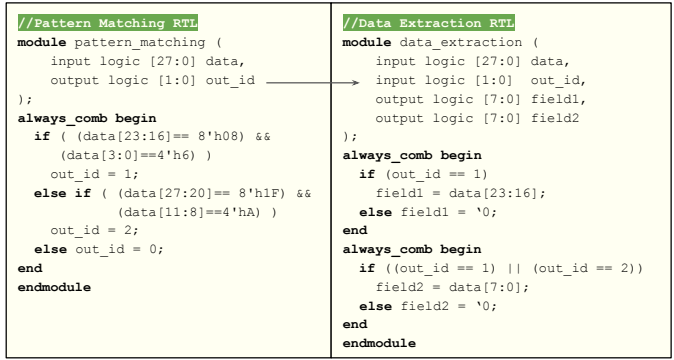

An example PIR with two pattern strings is shown in Fig.~\ref{fig:bit_mapping_constant}.
Each pattern has 7 nibbles (28 bits) containing constant hexadecimal values (\textcolor{matchblue}{\rule{1.2ex}{1.2ex}}) and wildcards (\textcolor{ignoregray}{\rule{1.2ex}{1.2ex}}), with output identifiers (\textcolor{green!20}{\rule{1.2ex}{1.2ex}}) set to \texttt{1} and \texttt{2}. The tuple (\textcolor{red!20}{\rule{1.2ex}{1.2ex}}) shows field extraction information per pattern.
The tool extracts the bit positions and length of each contiguous set of nibble characters and ignores the wildcards. 
Then, the template based RTL generator produces the pattern matching and data extraction implementations shown in Fig.~\ref{fig:bit_mapping_constant}b.
When the first pattern is matched, the two fields \texttt{data[23:16]} and \texttt{data[7:0]} are steered to the \texttt{field1} and \texttt{field2} output ports, respectively.
On the other hand, when the second pattern is matched, only \texttt{data[7:0]} is extracted and steered to the \texttt{field2} port.

\vspace{-4pt}
\subsection{Tool Configuration}
\label{sec:tool_config}

The backend takes a \texttt{json} configuration file as input, as shown in Listing~\ref{list:json_config}. This configuration enables the backend to: (1) decompose pattern matching into subproblems for optimized timing and resource utilization, and (2) generate logic for arbitrary datapath widths.

\begin{lstlisting}[linewidth=\columnwidth,breaklines=true,language=json,abovecaptionskip=-10pt, label={list:json_config}, caption={Tool configuration file example. \vspace{-0.5cm}}, style=topofpage]
  "|\color{keycolor}num\_groups|": 4,
  "|\color{keycolor}group\_datawidth|": 3,
  "|\color{keycolor}datapath\_width|": 6,
  "|\color{keycolor}invalid\_match\_output|": 0,
  "|\color{keycolor}custom\_tokens|": [{
      "|\color{keycolor}id|": "VLLL",
      "|\color{keycolor}type|": "port",
      "|\color{keycolor}prefix|": "myport",
      "|\color{keycolor}opcode|": "RANGE" }]
\end{lstlisting}

\subsubsection{Hierarchical Pattern Matching Parameters} 

The user can implement two-level hierarchical pattern matching by breaking pattern strings into \texttt{num\_groups} \emph{pattern substrings}, each \texttt{group\_datawidth} nibbles wide.
Each group is treated as an independent subproblem.
Repeated substrings within each group are deduplicated to reduce the size of the subproblem, and a \emph{pseudo output identifier} is assigned to each unique substring.
A second-stage module then maps the collected pseudo identifiers to final output identifiers.
This organization reduces comparator bits when repeated substrings are prevalent, improving resource utilization and timing.

\begin{figure}[t!]
\centering
\resizebox{\linewidth}{!}{%
  \begin{tikzpicture}[
    cell/.style={
        rectangle, 
        draw, 
        minimum width=0.6cm, 
        minimum height=0.6cm, 
        font=\ttfamily\scriptsize, 
        anchor=center,
        inner sep=0pt
    },
    header/.style={
        font=\bfseries\sffamily\small, 
        anchor=west
    },
    arrow_label/.style={
        font=\itshape\scriptsize, 
        midway, 
        right=0.3cm,
        align=left,
        color=black 
    },
    dotted_cell/.style={
        cell, 
        draw=black!50, 
        densely dashed, 
        fill=white
    },
]

\def\stepOneY{0}
\def\stepTwoY{-3} 
\def\stepThreeY{-7.2}

\def\colOneX{-1.8}
\def\colTwoX{1.2}
\def\colThreeX{4.2}
\def\colFourX{7.2}
\def\subRowH{0.7} 

\newcommand{\drawSubRow}[8]{
    \node[cell, fill=#6] at (#1, #2) {#3};
    \node[cell, fill=#6] at (#1+0.6, #2) {#4};
    \node[cell, fill=#6] at (#1+1.2, #2) {#5};
    \node[cell, fill=#8] at (#1+2, #2) {#7};
}
\newcommand{\drawDottedSubRow}[6]{
    \node[dotted_cell, fill=#6] at (#1, #2) {#3};
    \node[dotted_cell, fill=#6] at (#1+0.6, #2) {#4};
    \node[dotted_cell, fill=#6] at (#1+1.2, #2) {#5};
}

\newcommand{\drawGatherRow}[7]{
    \node[anchor=east, font=\sffamily\scriptsize] at (-0.2, #1) {#2};
    \node[cell, fill=ignoregray]   at (0.6, #1)     {#3};
    \node[cell, fill=green!20]  at (0.6+0.6, #1) {#4};
    \node[cell, fill=green!20]  at (0.6+1.2, #1) {#5};
    \node[cell, fill=ignoregray]   at (0.6+1.8, #1) {#6}; 
    \node[right=0.1cm, font=\scriptsize] at (2.8, #1) {$\rightarrow$ \textbf{#7}};
}


\node[anchor=east, font=\sffamily\scriptsize] at (-0.3, \stepOneY) {Pattern 1};
\foreach \char [count=\i] in {x,x,x,A,B,C,C,D,F,E,F,F} {
    \pgfmathtruncatemacro{\col}{\i-1}
    \ifnum\pdfstrcmp{\char}{x}=0
        \node[cell, fill=ignoregray] (p1_\i) at (\col*0.6, \stepOneY) {\char};
    \else
        \node[cell, fill=matchblue] (p1_\i) at (\col*0.6, \stepOneY) {\char};
    \fi
}
\node[cell, fill=green!20, right=0.3cm of p1_12] {1};

\def\pTwoY{\stepOneY-0.7}
\node[anchor=east, font=\sffamily\scriptsize] at (-0.3, \pTwoY) {Pattern 2};
\foreach \char [count=\i] in {A,B,C,C,D,E,F,E,A,x,x,x} {
    \pgfmathtruncatemacro{\col}{\i-1}
    \ifnum\pdfstrcmp{\char}{x}=0
        \node[cell, fill=ignoregray] (p2_\i) at (\col*0.6, \pTwoY) {\char};
    \else
        \node[cell, fill=matchblue] (p2_\i) at (\col*0.6, \pTwoY) {\char};
    \fi
}
\node[cell, fill=green!20, right=0.3cm of p2_12] {6};

\def\pThreeY{\pTwoY-0.7}
\node[anchor=east, font=\sffamily\scriptsize] at (-0.3, \pThreeY) {Pattern 3};
\foreach \char [count=\i] in {A,B,C,D,D,D,C,D,F,x,x,x} {
    \pgfmathtruncatemacro{\col}{\i-1}
    \ifnum\pdfstrcmp{\char}{x}=0
        \node[cell, fill=ignoregray] (p3_\i) at (\col*0.6, \pThreeY) {\char};
    \else
        \node[cell, fill=matchblue] (p3_\i) at (\col*0.6, \pThreeY) {\char};
    \fi
}
\node[cell, fill=green!20, right=0.3cm of p3_12] {7};

\draw[->, thick, black] (3.25, \pThreeY-0.4) -- (3.25, \stepTwoY+0.2) 
    node[arrow_label] {Deduplication \& Segmentation\\(4 columns, 3 characters each)};


\node[font=\bfseries\scriptsize] at (\colOneX+1.2, \stepTwoY) {Group 0};
\drawSubRow{\colOneX}{\stepTwoY-\subRowH}{x}{x}{x}{ignoregray}{x}{orange!20}
\drawSubRow{\colOneX}{\stepTwoY-2*\subRowH}{A}{B}{C}{matchblue}{1}{orange!20}
\drawDottedSubRow{\colOneX}{\stepTwoY-3*\subRowH}{A}{B}{C}{}
\draw [decorate, decoration={brace, amplitude=4pt, mirror}, thick, black]
    (\colOneX-0.3, \stepTwoY-3*\subRowH-0.4) -- (\colOneX+1.5, \stepTwoY-3*\subRowH-0.4)
    node [midway, below=4pt, font=\sffamily\tiny, align=center] {pattern substring};

\node[font=\bfseries\scriptsize] at (\colTwoX+1.2, \stepTwoY) {Group 1};
\drawSubRow{\colTwoX}{\stepTwoY-\subRowH}{A}{B}{C}{matchblue}{1}{orange!20}
\drawSubRow{\colTwoX}{\stepTwoY-2*\subRowH}{C}{D}{E}{matchblue}{2}{orange!20}
\drawSubRow{\colTwoX}{\stepTwoY-3*\subRowH}{D}{D}{D}{matchblue}{3}{orange!20}

\node[font=\bfseries\scriptsize] at (\colThreeX+1.2, \stepTwoY) {Group 2};
\drawSubRow{\colThreeX}{\stepTwoY-\subRowH}{C}{D}{F}{matchblue}{1}{orange!20}
\drawSubRow{\colThreeX}{\stepTwoY-2*\subRowH}{F}{E}{A}{matchblue}{2}{orange!20}
\drawDottedSubRow{\colThreeX}{\stepTwoY-3*\subRowH}{C}{D}{F}{}

\node[font=\bfseries\scriptsize] at (\colFourX+1.2, \stepTwoY) {Group 3};
\drawSubRow{\colFourX}{\stepTwoY-\subRowH}{E}{F}{F}{matchblue}{1}{orange!20}
\drawSubRow{\colFourX}{\stepTwoY-2*\subRowH}{x}{x}{x}{ignoregray}{x}{orange!20}
\drawDottedSubRow{\colFourX}{\stepTwoY-3*\subRowH}{x}{x}{x}{}

\draw[->, thick, black] (3.25, \stepTwoY-3.5*\subRowH-0.2) -- (3.25, \stepThreeY+0.5) 
    node[arrow_label] {Gather Pseudo IDs};


\node[anchor=east, font=\sffamily\scriptsize] at (0.3, \stepThreeY) {Pattern 1};
\node[cell, fill=ignoregray]   at (0.6, \stepThreeY)     {x};
\node[cell, fill=matchblue]  at (0.6+0.6, \stepThreeY) {1};
\node[cell, fill=matchblue]  at (0.6+1.2, \stepThreeY) {1};
\node[cell, fill=matchblue]  at (0.6+1.8, \stepThreeY) {1};
\node[cell, fill=green!20] at (0.6+2.7, \stepThreeY) {1};

\def\gTwoY{\stepThreeY-0.7}
\node[anchor=east, font=\sffamily\scriptsize] at (0.3, \gTwoY) {Pattern 2};
\node[cell, fill=matchblue]  at (0.6, \gTwoY)     {1};
\node[cell, fill=matchblue]  at (0.6+0.6, \gTwoY) {2};
\node[cell, fill=matchblue]  at (0.6+1.2, \gTwoY) {2};
\node[cell, fill=ignoregray]   at (0.6+1.8, \gTwoY) {x};
\node[cell, fill=green!20] at (0.6+2.7, \gTwoY) {6};

\def\gThreeY{\gTwoY-0.7}
\node[anchor=east, font=\sffamily\scriptsize] at (0.3, \gThreeY) {Pattern 3};
\node[cell, fill=matchblue]  at (0.6, \gThreeY)     {1};
\node[cell, fill=matchblue]  at (0.6+0.6, \gThreeY) {3};
\node[cell, fill=matchblue]  at (0.6+1.2, \gThreeY) {1};
\node[cell, fill=ignoregray]   at (0.6+1.8, \gThreeY) {x};
\node[cell, fill=green!20] at (0.6+2.7, \gThreeY) {7};

\end{tikzpicture}
}
\caption{Hierarchical pattern matching example with \texttt{num\_groups}=4 and \texttt{group\_datawidth}=3.}
\label{fig:h_pattern_matching}
\vspace{-10pt}
\end{figure}
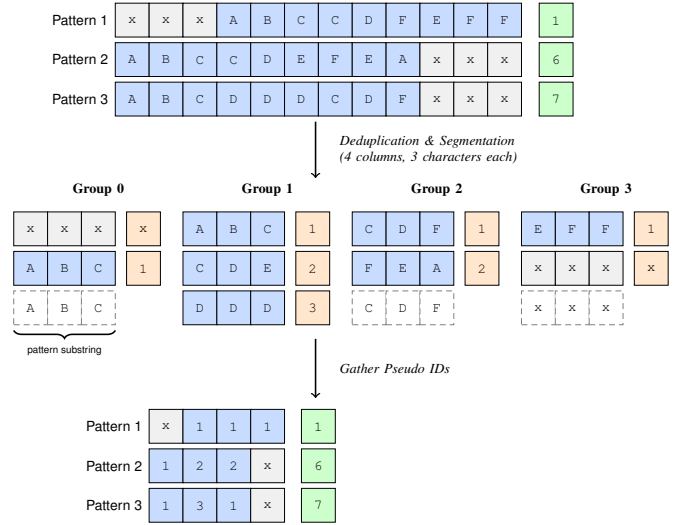

Fig.~\ref{fig:h_pattern_matching} shows an example with \texttt{num\_groups} and \texttt{group\_datawidth} set to 4 and 3, respectively.
Each of the 12-nibble pattern strings is split into four 3-nibble pattern substrings.
Following segmentation, the repeated pattern substrings \texttt{ABC}, \texttt{CDF}, and \texttt{xxx} are eliminated from groups 0, 2, and 3, respectively.
Then, a unique group-level pseudo output identifier (\textcolor{orange!20}{\rule{1.2ex}{1.2ex}}) is assigned to each pattern substring.
These pseudo output identifiers are used in a second-stage pattern matcher to produce the final output identifiers (\textcolor{green!20}{\rule{1.2ex}{1.2ex}}). 
Substrings containing do-not-care characters within a group can overlap. 
For instance, \texttt{6x} and \texttt{x3} both match input \texttt{63}, causing a collision. 
The tool automatically resolves such cases by further splitting the substrings until no collisions remain.

\subsubsection{Datapath Width Parameter} 
\label{subsubsec:multicycle}

\begin{figure}[t!]
    \centering
    \includegraphics[width=1\linewidth]{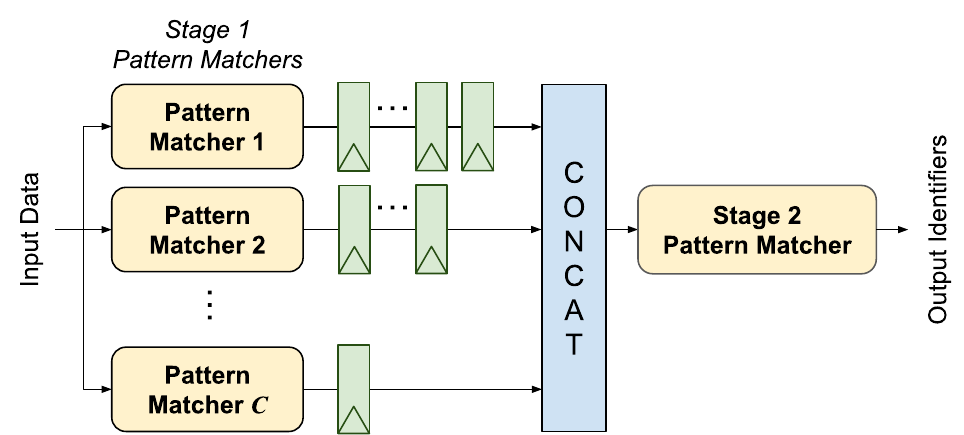}
    \caption{Generated pattern matching hardware for pattern strings straddling multiple cycles.}
    \label{fig:multicycle}
    \vspace{-10pt}
\end{figure}

The \texttt{datapath\_width} parameter specifies the pattern matching datapath width in nibbles and must satisfy:
\begin{equation*}
   \texttt{datapath\_width}  \bmod \texttt{group\_datawidth} = 0 
\end{equation*}
as each clock cycle must process integer number of complete groups. The number of cycles $C$ the full pattern string straddles is:
{
\small
\begin{equation*}
    C \times \text{\texttt{datapath\_width}} = \text{\texttt{num\_groups}} \times \text{\texttt{group\_datawidth}}
\end{equation*}
}

The tool generates $C$ distinct pattern matching modules processing flits in round-robin fashion.
Pseudo output identifiers from the $C$ modules are aligned using pipeline registers (see Fig.~\ref{fig:multicycle}) and concatenated as input to a second-stage module producing the final output identifiers.

The \texttt{invalid\_match\_output} parameter specifies the output identifier produced when the input data does not match any PIR pattern string. 
These knobs enable architectural design space exploration; however, determining optimal configurations currently requires slow RTL synthesis in the loop. 
Analytical resource estimation, without invoking a synthesis tool, remains an avenue for future work.

\vspace{-4pt}
\subsection{Custom Symbolic Token Parameters}
\label{sec:custom_tokens}

A key novel feature of our backend is extending the pattern matching alphabet beyond standard hexadecimal and wildcard characters via the \texttt{custom\_tokens} configuration parameter.
This parameter acts as a symbolic mapping, defining hardware generation rules for custom characters used in PIR pattern strings.
Each custom token definition consists of two components:

\begin{itemize}[leftmargin=*, itemindent=0pt, labelsep=3pt, align=left]
    \item \texttt{opcode}: Defines the logical predicate for comparison. The tool extends standard equality \texttt{(==)} to support inequality \texttt{(NE)}, set membership \texttt{(EQOR)}, and range validation \texttt{(RANGE)}. The schema is extensible, allowing users to add arbitrary hardware predicates (e.g., bit-counting or parity checks) with minimal backend modifications.
    \item \texttt{type}: Determines the operand source:
    \begin{itemize}
        \item \texttt{const}: Comparison operands are embedded directly in the configuration using the \texttt{values} parameter as immediate values in the generated RTL.
        \item \texttt{port}: The tool creates top-level input port(s) in the generated RTL, allowing match conditions to be driven by external system state, other modules, or software-configurable registers. Port naming is controlled via \texttt{prefix} and port count is determined by \texttt{opcode} (e.g., \texttt{RANGE} requires two ports).
    \end{itemize}
\end{itemize}

\begin{figure}[t!]
\centering
    \begin{subfigure}[b]{\linewidth} 
        \centering
        \begin{minipage}[b]{0.05\linewidth} 
            \centering
            {\footnotesize (a)}
        \end{minipage}
        \begin{minipage}[b]{0.6\linewidth} 
            \centering
            \resizebox{\linewidth}{!}{\begin{tikzpicture}[
    cell/.style={rectangle, draw, minimum width=0.7cm, minimum height=0.7cm, font=\ttfamily\large},
    bit_label/.style={font=\scriptsize\sffamily, color=gray!80!black},
    val_label/.style={font=\small\sffamily\bfseries}
]


    \node[cell, fill=orange] (c0) at (0,0) {V};
    \node[cell, fill=orange]  (c1) at (0.7,0) {L};
    \node[cell, fill=orange]  (c2) at (1.4,0) {L};
    \node[cell, fill=orange] (c3) at (2.1,0) {L};
    \node[cell, fill=matchblue] (c4) at (2.8,0) {7};
    \node[cell, fill=ignoregray] (c5) at (3.5,0) {x};
    \node[cell, fill=ignoregray]  (c6) at (4.2,0) {x};

    \node[cell, fill=green!20]  (c7) at (5.1,0) {5};

    \node[above=0.05cm of c0, bit_label] {27:24};
    \node[above=0.05cm of c6, bit_label] {3:0};

    \draw[<-, thick] (c0.north west) -- ++(0, 0.6) node[above, font=\sffamily\small] {MSB};
    \draw[<-, thick] (c6.north east) -- ++(0, 0.6) node[above, font=\sffamily\small] {LSB};

\end{tikzpicture}}
        \end{minipage}
        \label{subfig:bitmap2}
    \end{subfigure}\\[-2ex]
     \vspace{0.5cm} 

    \begin{subfigure}[b]{\linewidth}
        \centering
        \begin{minipage}[b]{0.05\linewidth}
            \centering
            {\footnotesize (b)}
            \vspace{1.5cm} 
        \end{minipage}
        \begin{minipage}[b]{0.6\linewidth}
            \centering
            \includegraphics[width=\linewidth]{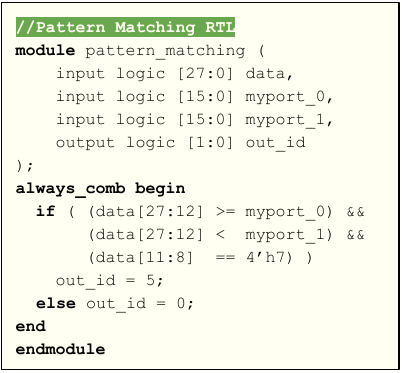}
        \end{minipage}
        \label{subfig:sv_pdf2}
    \end{subfigure}

\caption{%
(a) Bit map of a PIR entry with its output identifier shown in (\textcolor{green!20}{\rule{1.2ex}{1.2ex}}) and a custom token shown in (\textcolor{orange}{\rule{1.2ex}{1.2ex}}). (b) Generated RTL with range comparators for (\textcolor{orange}{\rule{1.2ex}{1.2ex}}) region, and equality comparator for (\textcolor{matchblue}{\rule{1.2ex}{1.2ex}}) region.
}
\label{fig:bit_mapping_custom}
\vspace{-10pt}
\end{figure}
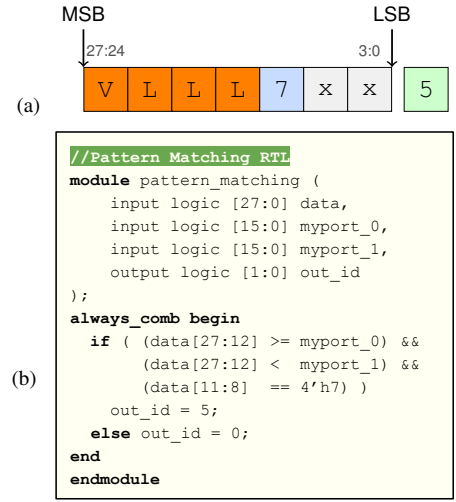

Fig.~\ref{fig:bit_mapping_custom} illustrates an example PIR entry containing the 4-nibble custom token defined in Listing~\ref{list:json_config}. 
The \texttt{custom} parameter maps the token \texttt{"VLLL"} to the \texttt{RANGE} opcode with source type \texttt{port}, instructing the backend to create two input ports and implement a magnitude comparator for the four nibbles specified by the custom token (i.e., \texttt{data[27:12]}). 
\vspace{-4pt}
\subsection{Custom Sparse Data Extraction Multiplexing} 
\label{subsec:custom_mux}
Traditional protocol parsers rely on generic barrel shifters for field extraction at runtime.
A generic barrel shifter selecting an $N$-bit field at any nibble-aligned offset within an $M$-bit datapath requires $N \times \lceil M/4\rceil$:1 multiplexers.
As datapath widths increase, these multiplexers cause significant area overhead, routing congestion, and timing degradation.
Our backend instead performs compile-time static analysis to generate sparse, customized multiplexing networks.
Because the frontend linearizes the parsing specification and assigns a unique output identifier to every possible pattern, the exact bit positions of every target field are known at generation time.
The extraction logic therefore generates a sparse multiplexer accounting only for the limited set of offsets at which each field can appear.
For example, extracting a 32-bit field from a 512-bit datapath would conventionally require 32$\times$512:1 multiplexers, but if the PIR analysis determines the field can only appear at 5 distinct offsets, the tool generates 32$\times$5:1 multiplexers, a significant reduction in area.

The following two sections demonstrate the tool's end-to-end capabilities through two case studies spanning distinct application domains: Ethernet protocol parsing (Section~\ref{sec:eth}) and network intrusion detection (Section~\ref{sec:snort}).
We then validate the impact of hierarchical decomposition on a controlled synthetic benchmark in Section~\ref{sec:decomposition}.
\section{Case Study I: Ethernet Protocol Parsing}
\vspace{-2pt}
\label{sec:eth}

We implement a P4-based frontend that compiles a standard P4 program into the PIR and tool configuration file consumed by our backend, demonstrating the end-to-end flow from a high-level protocol specification to synthesizable RTL.

\begin{lstlisting}[linewidth=\columnwidth,breaklines=true,language=C,abovecaptionskip=-10pt, label={list:p4_extract}, caption={P4 header declaration with extraction annotations. \vspace{-0.5cm}}, style=topofpage, escapeinside={(*@}{@*)}]
header udp_t {
    bit<16> srcPort__sp__;(*@\tikzmark{sp}@*)
    bit<16> dstPort__dp__;(*@\tikzmark{dp}@*)
    bit<16> length_;
    bit<16> checksum;
}
\end{lstlisting}
\begin{tikzpicture}[remember picture, overlay]
    \draw[->, thick, red!70!black] 
        ([xshift=4pt]pic cs:sp) -- ++(1.2cm,0) 
        node[right, font=\small\sffamily] {extracted as \texttt{sp}};
    \draw[->, thick, red!70!black] 
        ([xshift=4pt]pic cs:dp) -- ++(1.2cm,0) 
        node[right, font=\small\sffamily] {extracted as \texttt{dp}};
\end{tikzpicture}

\vspace{-4pt}
\subsection{Frontend P4 Compiler and Backend Integration}
The user defines parsing logic using a standard P4 program with header declarations, parser states, and \texttt{select()} transitions. 
To specify which fields should be extracted in generated RTL, we introduce a lightweight annotation convention.
Any header field whose name contains a double-underscore suffix \texttt{\_\_<id>\_\_} is marked for extraction, where \texttt{<id>} becomes the field identifier in the PIR. 
Fields without the suffix are used only for pattern matching and are not extracted. 
As shown in Listing~\ref{list:p4_extract}, declaring \texttt{srcPort\_\_sp\_\_} instructs the frontend to populate the optional extraction tuple in the PIR entry for that path, recording the field identifier, start bit position, and end bit position. 
The backend then uses this information to generate the data extraction RTL alongside the pattern matching logic, as described in Section~\ref{sec:tool_overview}. 
This convention keeps the P4 program fully compatible with standard P4 toolchains while providing our frontend with the extraction metadata it needs.

The frontend invokes the standard \texttt{p4c}~\cite{p4c} compiler with the 
BMV2~\cite{bmv2} backend to produce a \texttt{json} representation of the P4 program, encoding 
header types, field widths, and the parser state machine. It then traverses the parser's state graph via depth-first search to enumerate all valid protocol paths 
(e.g., Ethernet $\rightarrow$ IPv4 $\rightarrow$ TCP). Each root-to-leaf path is 
flattened into a nibble-granularity pattern string of hexadecimal constants and 
wildcards, assigned a unique output identifier, and annotated with 
\texttt{id:startpos:endpos} extraction tuples derived from \texttt{\_\_<id>\_\_} 
suffixed fields, forming the PIR. The frontend also generates a backend configuration 
file using two user-provided frontend parameters: \texttt{input\_width} and \texttt{pattern\_width}, 
setting backend \texttt{datapath\_width} and \texttt{group\_datawidth} equal to 
\texttt{input\_width} and computing backend \texttt{num\_groups} as 
\texttt{pattern\_width} / \texttt{input\_width}. The backend then consumes the PIR and configuration file to produce the pattern matching and data extraction RTL as detailed in Section~\ref{sec:tool_overview}.

\vspace{-4pt}
\subsection{Evaluation}
To evaluate the end-to-end flow, we generate two parser types consistent with prior work~\cite{gibb2013design, benacek2016p4, santiago2018p4}. 
We compare against these  academic baselines as they represent the latest publicly available parser RTL generators. 
Commercial P4 compilers (e.g., AMD Vitis Networking P4~\cite{amd2025vitisp4}) generate the parser, deparser, and match-action blocks as a single integrated pipeline, making it infeasible to isolate parser-only area and timing numbers for a fair comparison. 
Both parser types extract the standard five-tuple (protocol, source/destination IP, source/destination port) from the incoming packet stream:

\begin{itemize}[leftmargin=*, itemindent=0pt, labelsep=3pt, align=left]
\item \textit{Simple parser (SP):} Supports Ethernet, IPv4/IPv6 (with up to two extension headers), UDP, TCP, and ICMP/ICMPv6. The full pattern string length is 90 bytes.
\item \textit{Full parser (FP):} Extends SP with MPLS (up to two labels) and VLAN (up to two tags). The full pattern string length is 106 bytes.
\end{itemize}

\subsubsection{Comparison against prior work}

Table~\ref{tab:comprehensive_results} compares our generated designs against prior work on an AMD Virtex-7 (\texttt{xc7vx690tffg1761-2}), which is the target platform used in these prior works. 
To ensure a fair comparison, we match the exact datapath width used by each prior work. 
All designs were synthesized using Vivado 2025.2 in out-of-context (OOC) mode with default settings. 
The synthesized RTL uses fully registered inputs and outputs with valid-ready flow control. 
We sourced the resource utilization numbers for prior works from~\cite{santiago2018p4}, which consolidates results from earlier works. For Gain/Loss calculations, we consider the \textit{best} result from prior work, ensuring our reported gains are conservative.

\begin{table}[t]
\centering
\caption{Comparison of Parser Results across Benchmarks (N/R = Not Reported; N/C = Not Calculated)}
\label{tab:comprehensive_results}
\setlength{\tabcolsep}{4pt} 
\begin{tabularx}{\columnwidth}{c l YYYYY} 

\toprule
& \textbf{Work} & \textbf{Freq} & \textbf{T-put} & \textbf{Latency} & \textbf{LUTs} & \textbf{FFs} \\
& & [MHz] & [Gb/s] & [ns] & & \\
\midrule
\multirow{3}{*}{\rotatebox[origin=c]{90}{\parbox{1.1cm}{\centering\small\textbf{SP\\(256b)}}}}

& \cite{gibb2013design} & 178.6 & 46 & N/R & 6865 & 1851 \\
\cline{2-7}
& \cellcolor{MyWorkColor} ours & \cellcolor{MyWorkColor} {583} & \cellcolor{MyWorkColor} {149.2} & \cellcolor{MyWorkColor} {15.4} & \cellcolor{MyWorkColor} {179} & \cellcolor{MyWorkColor} {2116} \\
\cline{2-7}
& \textit{Gain/Loss \%} & \textit{+226\%} & \textit{+224\%} & \textit{N/C} & \textit{-97\%} & \textit{14\%} \\
\specialrule{1.5pt}{1pt}{1pt} 
\multirow{4}{*}{\rotatebox[origin=c]{90}{\parbox{1.4cm}{\centering\small\textbf{SP\\(320b)}}}}

& \cite{santiago2018p4} & 312.5 & 100 & 19.2 & 4270 & 6163 \\
& Hybrid \cite{santiago2018p4} and \cite{benacek2016p4}  & 312.5 & 100 & 28.8 & 4699 & 7254 \\

& ours & {562} & {179.8} & {16} & {180} & {2564} \\
\cline{2-7}
& \textit{Gain/Loss \%} & \textit{+80\%} & \textit{+80\%} & \textit{-17\%} & \textit{-96\%} & \textit{-58\%} \\
\specialrule{1.5pt}{1pt}{1pt} 
\multirow{3}{*}{\rotatebox[origin=c]{90}{\parbox{1.1cm}{\centering\small\textbf{FP\\(64b)}}}}

& \cite{gibb2013design} & 172.2 & 11 & N/R & 3789 & 1425 \\

& ours & {379} & {24.3} & {81.8} & {1331} & {2240} \\
\cline{2-7}
& \textit{Gain/Loss \%} & \textit{+120\%} & \textit{+121\%} & \textit{N/C} & \textit{-65\%} & \textit{57\%} \\
\specialrule{1.5pt}{1pt}{1pt} 
\multirow{4}{*}{\rotatebox[origin=c]{90}{\parbox{1.4cm}{\centering\small\textbf{FP\\(320b)}}}}

& \cite{santiago2018p4} & 312.5 & 100 & 25.6 & 6046 & 8900 \\
& Hybrid \cite{santiago2018p4} and \cite{benacek2016p4}  & 312.5 & 100 & 41.6 & 6450 & 10308 \\

& ours & {359} & {114.9} & {25.1} & {1357} & {2619} \\
\cline{2-7}
& \textit{Gain/Loss \%} & \textit{+15\%} & \textit{+15\%} & \textit{-2\%} & \textit{-78\%} & \textit{-71\%} \\
\specialrule{1.5pt}{1pt}{1pt} 
\multirow{3}{*}{\rotatebox[origin=c]{90}{\parbox{1.1cm}{\centering\small\textbf{FP\\(512b)}}}}

& \cite{benacek2016p4} & 195.3 & 100 & 46.1 & 10103 & 5537 \\

& ours & {379} & {194} & {18.5} & {1352} & {2938} \\
\cline{2-7}
& \textit{Gain/Loss \%} & \textit{+94\%} & \textit{+94\%} & \textit{-60\%} & \textit{-87\%} & \textit{-47\%} \\
\bottomrule
\end{tabularx}
\vspace{-10pt}
\end{table}

Our framework pipelines the generated designs heavily and applies \texttt{(* DONT\_TOUCH = TRUE *)} on all FFs to avoid tool retiming. 
This aggressive pipelining leads to a higher FF count compared to prior work in some cases. 
Our generated designs use 65--97\% less LUTs, while the FF utilization varies with datapath width; wider datapaths use 47--71\% less FFs, while the narrowest configurations (\textit{SP}(256b) and \textit{FP}(64b)) use more FFs than prior work (+14\% and +57\%, respectively). 
For a fixed pattern length, the datapath width determines the number of groups (a narrower datapath splits the pattern across more groups), and each group adds a pipeline stage, so more groups use more pipeline registers. 
At low datapath widths this group count is highest, and combined with our aggressive pipelining, this inflates the FF count.
As for Fmax and throughput, \textit{FP}(320b) shows the smallest gain (+15\%), as prior work already operated at 312.5~MHz. 
All other benchmarks achieve gains of 80\% or more, reaching +226\% for \textit{SP}(256b) where prior work ran at only 178.6~MHz. 
Across all benchmarks, our latency is lower than or equal to prior work, with the largest improvement at \textit{FP}(512b) where we reduce end-to-end latency by 60\% (from 46.1~ns to 18.5~ns).

\subsubsection{Scaling across datapath widths}
\begin{figure}[t!]
  \centering
  \includegraphics[width=0.5\textwidth]{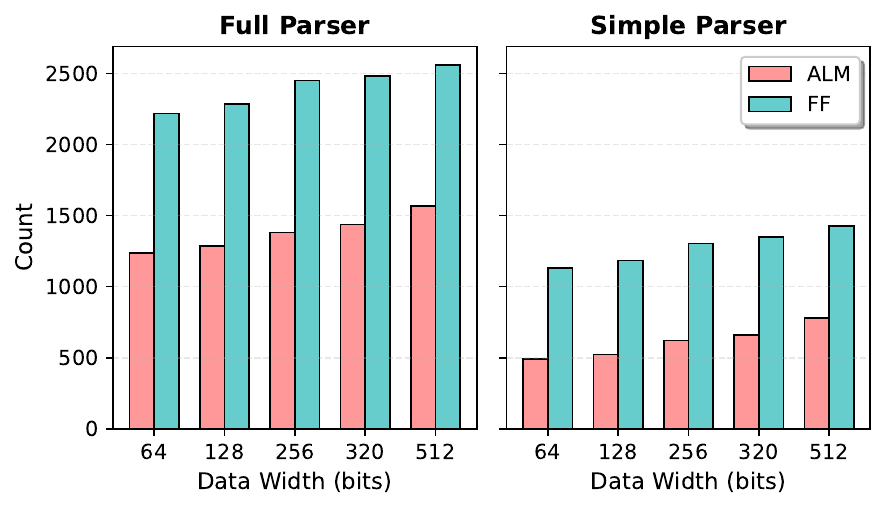}
  \caption{ALM/FF utilization vs. datapath width for Altera Agilex 5 FPGAs.}
  \label{fig:intel_alm_vs_dp_unique}
  \vspace{-10pt}
\end{figure}

\begin{figure}[t!]
  \centering
  \includegraphics[width=0.5\textwidth]{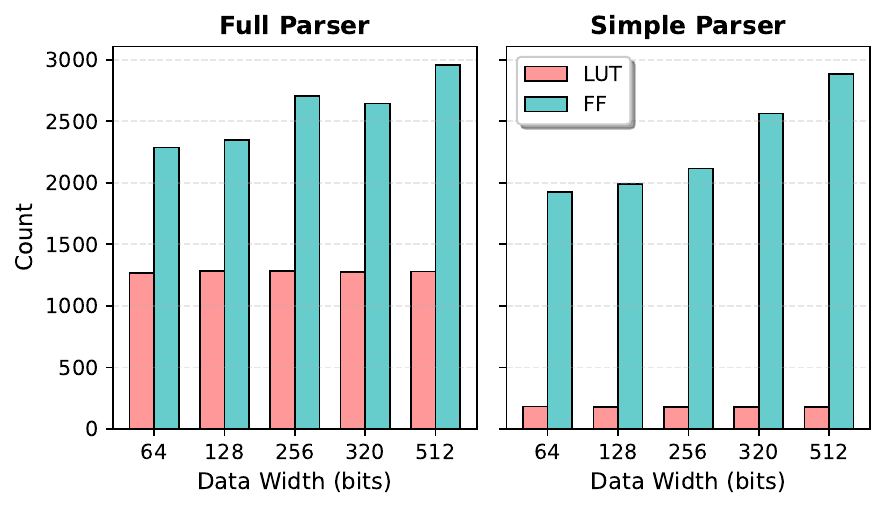}
  \caption{LUT/FF utilization vs. datapath width for AMD FPGAs.}
  \label{fig:xil_lut_vs_dp_unique}
  \vspace{-10pt}
\end{figure}

To characterize how our designs scale beyond the datapath widths used in prior work, we sweep \texttt{datapath\_width} from 64 to 512 bits on two modern FPGAs using the highest speed grade available from each vendor: Altera Agilex 5 (\texttt{A5DD064AB32BI1V}) and AMD Versal (\texttt{xcvc1902-vsva2197-3HP-e-S}). Results are shown in Fig.~\ref{fig:intel_alm_vs_dp_unique} and Fig.~\ref{fig:xil_lut_vs_dp_unique}.

Both parsers exhibit sub-linear LUT/ALM growth as the datapath widens. 
For a fixed pattern size, the datapath width determines how the pattern is partitioned into groups, setting \texttt{num\_groups} and \texttt{group\_datawidth}. 
A wider datapath has more bits per group, and therefore wider pipeline registers are used resulting in higher FF utilization as datapath width increases.
Also, wider groups require larger field comparators, resulting in an increase in logic utilization which is more noticeable on the Agilex 5 device. 

\section{Case Study II: Snort Rule Hardware Mapping}
\vspace{-2pt}
\label{sec:snort}

This case study demonstrates the versatility of our parser generation backend by applying it to network intrusion detection.
Rather than parsing protocol headers, the backend is used to generate hardware rule matchers for Snort~\cite{Snort3}, the de facto standard for network intrusion detection and prevention.
Crucially, this case study requires no modifications to the backend; only a different frontend is developed, validating that our PIR and custom symbolic tokens generalize beyond conventional protocol parsing.

Snort rules specify packet matching conditions using header fields (IPs, ports, protocol) combined with payload
inspection (content strings at specific offsets).
Many rules employ features such as Classless Inter-Domain Routing (CIDR) blocks for network ranges, comma-separated
IP lists, negation operators, and port ranges, operations that go beyond simple equality matching.
Existing hardware parser generators, which are designed around P4's equality-based match semantics, cannot express
these operations.
Our backend's custom symbolic tokens (\texttt{RANGE}, \texttt{EQOR}, \texttt{NE}) map directly to these Snort constructs, enabling hardware rule matching that would otherwise require hand-crafted RTL.

Systems such as Pigasus~\cite{zhao2020achieving} target full intrusion detection and prevention system (IDS/IPS) pipelines at 100~Gbps, implementing TCP reassembly, multi-string pattern matching, and offloading complex features like Perl Compatible Regular Expression (PCRE) to the CPU.
Our tool targets a complementary point in this design space: lightweight, full hardware implementation of rule matchers for header fields and simple content inspection suitable for first-stage packet filtering, protocol violation detection, and IP-based blocklisting.
In such cases, line-rate performance is critical and full stateful inspection can be handled downstream or offloaded if needed. 
Table~\ref{tab:snort_features} summarizes the Snort rule features supported by our frontend and their mapping to backend operators.
We do not support PCRE regular expressions or stateful flow tracking, which are outside the scope of static pattern matching.

\begin{table}[t!]
\centering
\caption{Snort Rule Features and Hardware Mapping}
\label{tab:snort_features}
\begin{tabular}{lcc}
\toprule
\textbf{Feature} & \textbf{Supported} & \textbf{Backend Operator} \\
\midrule
IP exact match & $\checkmark$ & equality \\
IP list (comma-separated) & $\checkmark$ & \texttt{EQOR} \\
CIDR blocks & $\checkmark$ & \texttt{RANGE} \\
IP ranges & $\checkmark$ & \texttt{RANGE} \\
Negation (\texttt{!}) & $\checkmark$ & \texttt{NE} \\
Port ranges & $\checkmark$ & \texttt{RANGE} \\
Content string & $\checkmark$ & equality \\
Content offset/depth & $\checkmark$ & rule expansion \\
PCRE (regex) & $\times$ & --- \\
Flow/stream state & $\times$ & --- \\
\bottomrule
\end{tabular}
\vspace{-10pt}
\end{table}

\subsection{Frontend Compiler}
The Snort frontend compiler parses rule strings from a text file and maps them to PIR entries through the following steps:

\begin{itemize}[leftmargin=*, itemindent=0pt, labelsep=3pt, align=left]
    \item \textit{Packet Pattern Construction:} The compiler constructs an Ethernet/IPv4/TCP (or UDP/ICMP) packet pattern from the rule header, placing specified field values (source/destination IPs, ports, protocol) at their correct byte offsets and marking all unspecified fields as wildcards (\texttt{x}).

    \item \textit{Custom Token Assignment:} Fields requiring non-equality comparisons are assigned custom symbolic tokens in the pattern string, with corresponding entries generated in the backend \texttt{json} configuration file.

    \item \textit{Content Positioning and Rule Expansion:} When a rule specifies \texttt{content} with \texttt{offset} $o$ and \texttt{depth} $d$, a content string of length $k$ can appear at any byte position in $[o, d-k]$. The frontend generates $(d - o - k + 1)$ separate PIR entries, one for each valid starting position. This compile-time expansion converts sliding-window searches into parallel static pattern matches, trading hardware area for throughput.
    
    \item \textit{Action Encoding:} The rule action (e.g., \texttt{alert}, \texttt{drop}), rule index, and message identifier are encoded together as a compact binary literal and assigned as the PIR \texttt{output\_id}. This enables the generated hardware to report which rule matched and what action to take without additional lookups.
\end{itemize}

\vspace{-4pt}
\subsection{Compilation Example}
\begin{figure}[t!]
  \centering
  \includegraphics[width=0.5\textwidth]{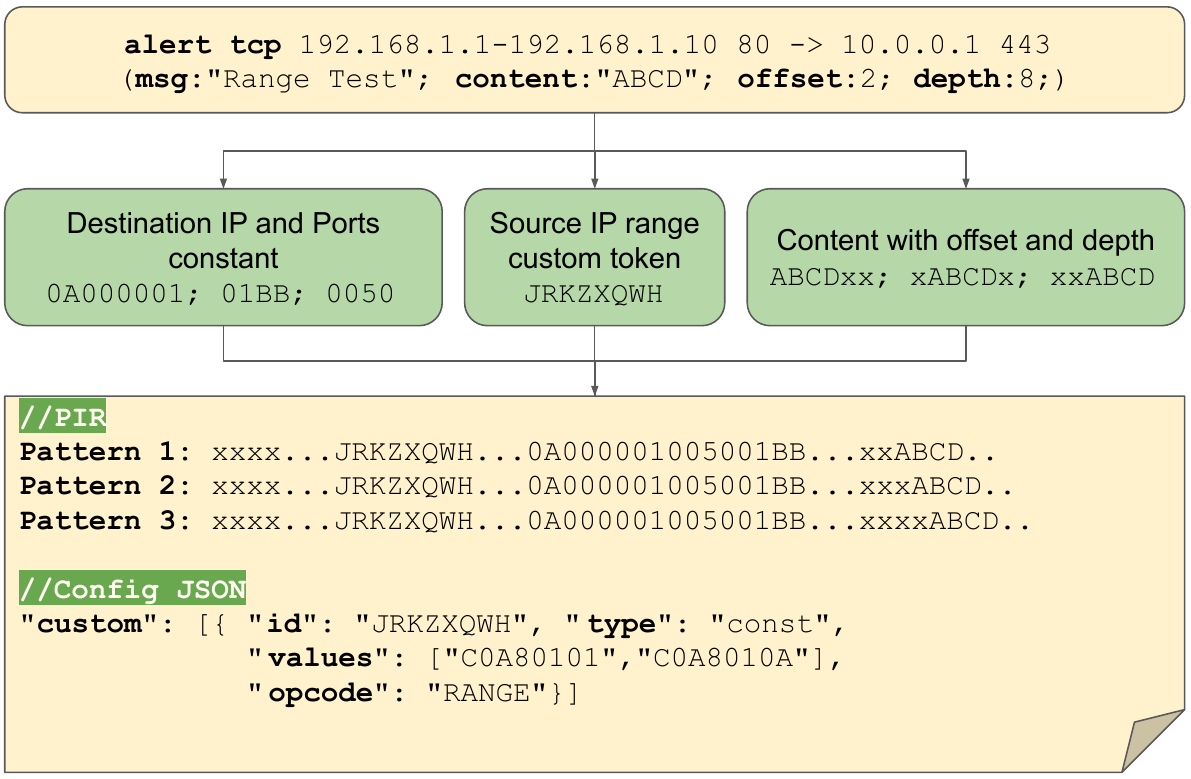}
  \caption{Snort rule compilation flow: the frontend maps header fields to hex constants and custom tokens, then expands the content constraint into multiple PIR entries. The backend configuration maps the custom token to a \texttt{RANGE} comparator with the specified IP bounds.}
  \label{fig:snort_example}
  \vspace{-5pt}
\end{figure}

Fig.~\ref{fig:snort_example} illustrates the complete compilation flow for a rule that triggers an \texttt{alert} for TCP packets from source IPs in the range 192.168.1.1--192.168.1.10 on port 80, destined for 10.0.0.1 on port 443, containing \texttt{"ABCD"} within byte offsets 2--8 of the payload.

The source IP range cannot be encoded as a fixed hex value in the PIR, so the compiler assigns a placeholder custom random token (\texttt{JRKZXQWH}) and generates a \texttt{RANGE} opcode entry in the configuration with bounds \texttt{192.168.1.1(0xC0A80101)}--\texttt{192.168.1.10(0xC0A8010A)}.

The destination IP and L4 ports are encoded directly as hex constants.
The content constraint \texttt{"ABCD"} with \texttt{offset:2} and \texttt{depth:8} allows the 4-byte string to start at payload offsets 2, 3, or 4, so the compiler generates three separate pattern entries, each identical except for the payload offset where \texttt{ABCD} appears.

\vspace{-4pt}
\subsection{Evaluation}

To evaluate the Snort frontend, we compile a benchmark of 10 rules that collectively exercise all supported features: exact IP matching, CIDR blocks, IP lists, IP ranges, negation, port ranges, and content with offset/depth expansion. 
As noted above, our goal in this case study is not to implement a production-grade IDS/IPS. 
This 10-rule benchmark instead showcases that the same backend, reusing the same PIR and custom symbolic tokens without any modifications, can be coupled with a non-P4 frontend to generate functionally correct, synthesizable, and efficient hardware for a domain beyond protocol parsing, across all supported feature types.

\begin{table}[t!]
\centering
\caption{Snort rule matcher synthesis results (10 rules).}
\label{tab:snort_synth}
\begin{tabular}{lccc}
\toprule
\textbf{Device} & \textbf{Fmax [MHz]} & \textbf{LUTs/ALMs} & \textbf{FFs} \\
\midrule
AMD Versal & 714 & 234 & 97 \\
Altera Agilex 5 & 712 & 706 & 99 \\
\bottomrule
\end{tabular}
\vspace{-10pt}
\end{table}

Table~\ref{tab:snort_synth} presents the synthesis results on a 1024-bit datapath for a 10-rule benchmark designed
to exercise all supported feature types (Table~\ref{tab:snort_features}).
Note that rules using content matching with offset/depth windows expand into multiple PIR entries (one per valid starting position), so area grows with both the number of rules and the content expansion factor.
\section{Impact of Hierarchical Decomposition}
\vspace{-2pt}
\label{sec:decomposition}

\begin{figure*}[t!]
    \centering
    \includegraphics[width=1\linewidth]{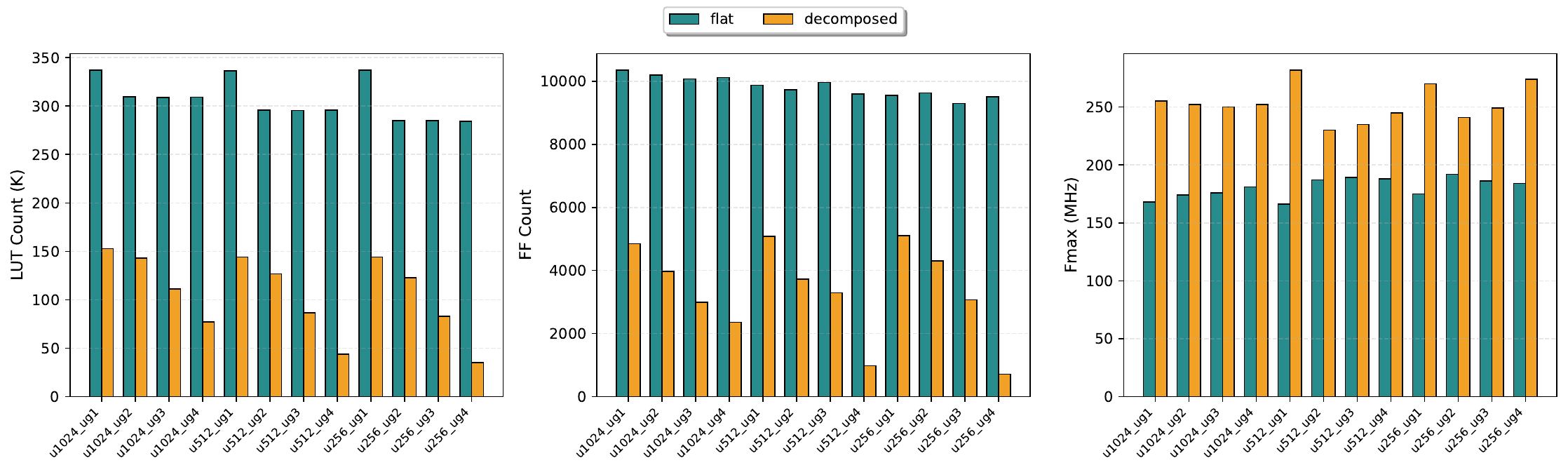}
    \caption{Comparison of synthetic benchmarks highlighting the advantages of pattern decomposition in terms of LUT count, FF count, and frequency on an AMD FPGA.}
    \label{fig:synthetic_decomposition}
    \vspace{-12pt}
\end{figure*}

Hierarchical decomposition is a key feature of our tool. 
It enables pattern matching across arbitrary datapath widths and assists vendor tool compilers by breaking a monolithic matching problem into smaller, independently optimizable subproblems. 
The Ethernet parsers evaluated in Section~\ref{sec:eth} use this decomposition, and the results in Table~\ref{tab:comprehensive_results} reflect its benefit. 
To isolate and quantify the impact of decomposition independently of protocol-specific effects, we design a controlled synthetic benchmark suite with varying levels of substring repetition. 
All patterns consist solely of hexadecimal characters with no wildcards or special tokens. 
We generate 4,096 pattern strings each 80 bits wide and assign a unique output identifier to each, effectively constructing an 80-bit input, $\lceil\log_2(4096+1)\rceil$-bit output lookup table (+1 for an invalid match output identifier). 
The pattern count and width are chosen arbitrarily to demonstrate how decomposing the matching problem into subproblems results in producing faster and more resource-efficient designs.

For benchmark construction, we treat the dataset as a 4096$\times$80 matrix. We partition this into four 4096$\times$20 groups. This allows us to independently control the repetition characteristics within each group. The benchmark suite sweeps two independent dimensions of repetition across these groups:

\begin{itemize}[leftmargin=*, itemindent=0pt, labelsep=3pt, align=left]
    \item \textit{Uniqueness per group (u):} The number of distinct 20-bit substrings within a group, drawn from \{256, 512, 1024\}, corresponding to each unique substring appearing 16$\times$, 8$\times$, and 4$\times$ across the 4,096 patterns, respectively.
    \item \textit{Number of controlled groups (ug):} Specifies how many of the four groups are assigned a controlled \textit{u}.
\end{itemize}

Crossing these two dimensions yields $12$ configurations.
For each configuration, we generate two functionally identical RTL variants: a \textit{flat} design (\texttt{num\_groups} = 1) that treats each 80-bit pattern as a single monolithic unit, ignoring the group structure entirely, and a \textit{decomposed} design (\texttt{num\_groups} = 4, \texttt{group\_datawidth} = 5) that aligns its partitioning with the benchmark groups, solving each 20-bit substring independently. The expectation is that as repetition increases, either through lower \textit{u} or higher \textit{ug}, the \textit{decomposed} design should show greater improvement in terms of area and speed over the \textit{flat} design, since more repetition within groups translates directly into more sharing that the decomposed subproblems can exploit.

\vspace{-4pt}
\subsection{Results}

Fig.~\ref{fig:synthetic_decomposition} reports Fmax, LUT, and FF for all 12 configurations, comparing the \textit{flat} and \textit{decomposed} variants across the full repetition sweep on an AMD Versal (\texttt{xcvc1902-vsva2197-3HP-e-S}) using Vivado 2025.2. 
As in our other experiments, we mark all FFs with \texttt{dont\_touch} synthesis directives such that Vivado does not retime pipeline registers, which otherwise distorts FF counts without improving Fmax.

\subsubsection{LUT utilization} 
The \textit{flat} design exposes a fundamental limitation: it is largely insensitive to substring repetition. For a fixed \textit{u}, its LUT count drops slightly from \textit{ug=1} to \textit{ug=2} and then plateaus (e.g., at \textit{u=256} it holds at $\sim$285K LUTs for \textit{ug}=2,3,4). 
Because the full 80-bit pattern is presented as a single monolithic unit, the vendor compiler cannot discover or share the repeated substrings.

The \textit{decomposed} design solves each 20-bit group independently and therefore exploits this repetition directly, with gains that scale with how much repetition is exposed. Even at \textit{ug=1},
where almost no repetition is controlled, splitting the problem into smaller subproblems already yields a $\sim$2.3$\times$ LUT reduction from structure alone. As repetition increases (higher
\textit{ug} or lower \textit{u}) the reduction compounds, reaching 6.7$\times$ at \textit{u=512, ug=4} (296K vs.\ 44K LUTs) and a peak of 8$\times$ at \textit{u=256, ug=4} (284K vs.\ 35K LUTs).

\subsubsection{Fmax} 
Decomposition also raises the clock frequency across every configuration, from the \textit{flat} design's 166--192~MHz to the \textit{decomposed} design's 230--282~MHz, roughly a 1.5$\times$ improvement. 
The gain comes from partitioning the 80-bit match into four independent 20-bit subproblems, which shortens the combinational critical path regardless of the repetition structure.

\subsubsection{FF utilization} 
The same partitioning also reduces FF utilization. 
The \textit{flat} design uses $\sim$9.3--10.4K FFs independent of repetition, whereas the \textit{decomposed} design uses far fewer FFs that are reduced from $\sim$5.1K FFs at \textit{ug=1} down to 715 FFs at \textit{u=256, ug=4} as repetition increases.

  Overall, these results confirm that vendor compilers cannot exploit substring repetition in \textit{flat} designs. Aligning the decomposition with the repetition structure yields up to 8$\times$ fewer
  LUTs and ${\sim}1.5\times$ higher Fmax, with the benefit growing directly with the degree of repetition.
\section{Limitations and Future Work}
\vspace{-2pt}
\label{sec:limitations}

  While our tool demonstrates strong results across both case studies, several limitations present opportunities for  
  future improvement.
  Custom symbolic tokens cannot be split across group boundaries during hierarchical decomposition; a token must reside entirely within a single group, constraining \texttt{num\_groups} and \texttt{group\_datawidth} for patterns
  with wide tokens.
  Additionally, the token system currently supports a fixed set of predicates (\texttt{RANGE}, \texttt{NE},
  \texttt{EQOR}).
  Extending this to operations such as bit counting, parity checks, or masked comparisons, and relaxing the
  cross-group constraint, would broaden applicability to domains beyond networking.
\section{Conclusion}
\vspace{-2pt}
\label{sec:conclusion}

This paper presented a versatile open-source tool for automatic hardware parser generation, built on a parsing intermediate representation (PIR) that decouples application-specific frontends from an optimized RTL generation backend. The custom symbolic token system extends pattern matching beyond equality to support range validation, negation, and external port comparisons, operations that no prior hardware parser generator can express natively. Two case studies validated the tool's generality: an Ethernet protocol parsing frontend that outperforms prior academic baselines across
 most reported metrics, and a Snort intrusion detection frontend that generates efficient hardware rule matchers using the
 same unmodified backend. A controlled synthetic study further confirmed that hierarchical decomposition, by breaking
 monolithic matching into independently optimizable subproblems, yields substantial area and frequency improvements that
 vendor tool compilers cannot achieve on flat designs. Unlike commercial P4 compilers that produce encrypted,
 vendor-locked IP, our template-driven approach generates structured, human-readable, and vendor-agnostic RTL. The tool is
  open-sourced to enable rapid hardware acceleration across diverse pattern-matching domains.

\bibliographystyle{IEEEtran}
\bibliography{references}

\end{document}